\documentclass[12pt]{article}
\usepackage{graphicx}
\usepackage{epsfig}
\usepackage{latexsym}
\usepackage{latexsym}
\usepackage{amssymb}
\usepackage{cite}
\newcommand{\labell}[1]{\label{#1}}
\newcommand{\reef}[1]{(\ref{#1})}

\newcommand{\bibbyitem}[1]{\bibitem{#1}}

\def\via{{\it via}}

\def\eg{{\it e.g.}}

\DeclareSymbolFont{AMSb}{U}{msb}{m}{n}
\DeclareMathSymbol{\IN}{\mathbin}{AMSb}{"4E}
\DeclareMathSymbol{\IZ}{\mathbin}{AMSb}{"5A}
\DeclareMathSymbol{\IR}{\mathbin}{AMSb}{"52}
\DeclareMathSymbol{\Q}{\mathbin}{AMSb}{"51}
\DeclareMathSymbol{\II}{\mathbin}{AMSb}{"49}
\DeclareMathSymbol{\IC}{\mathbin}{AMSb}{"43}
\DeclareMathSymbol{\IP}{\mathbin}{AMSb}{"50}
\DeclareMathSymbol{\IH}{\mathbin}{AMSb}{"48}
\DeclareMathSymbol\IA{\mathalpha}{AMSb}{"41}
\DeclareMathSymbol\IS{\mathalpha}{AMSb}{"53}

\def\Q{{\cal Q}}

\begin{document}


\hbox{$\phantom{.}$}

\bigskip
\bigskip
\begin{center}
   {\Large \bf  Holographic RG Flows and Universal Structures}

\bigskip

{\Large \bf on the Coulomb Branch of}

\bigskip

{\Large  \bf ${\cal N}{=}2$ Supersymmetric
Large $N$  Gauge Theory}
   \end{center}

\bigskip
\bigskip
\bigskip
\bigskip

\centerline{\bf James E. Carlisle, Clifford V. Johnson}

\bigskip
\bigskip
\bigskip

\centerline{\it Centre
for Particle Theory}
  \centerline{\it Department of Mathematical Sciences}
\centerline{\it University of
Durham}
\centerline{\it Durham, DH1 3LE, U.K.}

\centerline{$\phantom{and}$}

\bigskip

\centerline{\small \tt
j.e.carlisle@durham.ac.uk,
 c.v.johnson@durham.ac.uk}

\bigskip
\bigskip
\bigskip


\begin{abstract}
  We report on our results of D3--brane probing a large class of
  generalised type~IIB supergravity solutions presented very recently
  in the literature.  The structure of the solutions is controlled by
  a single non--linear differential equation. These solutions
  correspond to renormalisation group flows from pure ${\cal N}{=}4$
  supersymmetric gauge theory to an ${\cal N}{=}2$ gauge theory with a
  massive adjoint scalar. The gauge group is $SU(N)$ with $N$ large.
  After presenting the general result, we focus on one of the new
  solutions, solving for the specific coordinates needed to display
  the explicit metric on the moduli space. We obtain an appropriately
  holomorphic result for the coupling. We look for the singular locus,
  and interestingly, the final result again manifests itself in terms
  of a square root branch cut on the complex plane, as previously
  found for a set of solutions for which the details are very
  different. This, together with the existence of the single simple
  non--linear differential equation, is further evidence in support of
  an earlier suggestion that there is a very simple model ---perhaps a
  matrix model with relation to the Calogero--Moser integrable
  system--- underlying this gauge theory physics.

\end{abstract}
\newpage \baselineskip=18pt \setcounter{footnote}{0}

\section{Introductory Remarks}
\label{sec:introduction}
Last week, 
a new solution to type~IIB supergravity was presented in
ref.\cite{Pilch:2003jg}. It beautifully clarified and considerably
generalised the structure of an already interesting family of
supergravity solutions presented in ref.\cite{Pilch:2000ue} (see also
ref.\cite{freed1}.) Those earlier supergravity solutions are
asymptotic to the maximally supersymmetric AdS$_5\times S^5$ solution,
and \via\ the AdS/CFT correspondence\cite{adscft,adscft2}, and
generalisations thereof, represent ``Holographic Renormalisation Group
Flow''\cite{gppz1,dz1} of the four dimensional ${\cal N}{=}4$
supersymmetric pure Yang--Mills theory to ${\cal N}{=}2$ supersymmetric
gauge theory with a massive hypermultiplet in the adjoint
representation of $SU(N)$, in the large $N$ limit.  These new, more
general solutions preserve the same supersymmetries, and are also
believed to represent the same type of physics.

In this short note we carry out a study in the spirit of
refs.\cite{Buchel:2000cn,Evans:2000ct}. There were puzzling
singularities in the supergravity solutions of
ref.\cite{Pilch:2000ue} obscuring the gauge theory interpretation
considerably\cite{Pilch:2000ue,freed1}. The idea was to probe the
geometry with the most natural object to hand ---one of the
constituent D3--branes--- in an effort to determine the correct
physics. This had borne considerable fruit in a study reported in
ref.\cite{Johnson:1999qt}, where the nature of the singular behaviour
was understood, unphysical singularities were removed, and the new
phenomenon called the ``enhan\c con mechanism'' proved to be well adapted
to the task of clarifying the physics. As the situation in
ref.\cite{Pilch:2000ue} preserved the same supersymmetries as the
geometries under discussion in ref.\cite{Johnson:1999qt} (where in fact
the motivation was also to find gauge duals of ${\cal N}{=}2$ four
dimensional gauge theory), it was very natural to bring the same tools
to understanding the geometries of ref.\cite{Pilch:2000ue}.
Those studies were successful, and in light of the existence of the
more general class of solutions presented recently in
ref.\cite{Pilch:2003jg}, it is natural to carry out the same study
here. We probe the geometries and derive the effective Lagrangian for
the general form  of solutions in section~\ref{sec:proberesult}.

The detailed form of a solution is seeded by a non--linear
differential equation which yields a single
function\cite{Pilch:2003jg}. This is a difficult equation, and so far
only two families of solution are known. The first is the previously
known family\cite{Pilch:2000ue,freed1}. In
section~\ref{sec:specialize}, as a review and for contrast, we
specialise our result to that case, and recover the results of
ref.\cite{Buchel:2000cn,Evans:2000ct}. We particularly follow the
lines of ref.\cite{Buchel:2000cn} in moving away from the natural
supergravity coordinates and finding new coordinates on the moduli
space that are adapted to the full discussion of the low energy
effective Lagrangian of the ${\cal N}{=}2$ gauge theory. In particular,
in the spirit of ref.\cite{Seiberg:1994rs},
ref.\cite{Buchel:2000cn} exhibited explicit holomorphy and identified
in the new coordinates the locations of the points 
where there are singularities in the gauge coupling.

In section~\ref{anotherexample} we carry out a similar analysis for
the second family of exact solutions to the differential equation
exhibited in ref.\cite{Pilch:2003jg}. Whilst the solution is very
different from the earlier one, we find that once we have identified
the natural coordinates, the locus of singular points is controlled by
a very similar functional dependence as in the previous example: there is
a dense locus of singular points on a straight line segment controlled
by an (inverse) square root branch cut in the complex coupling $\tau$.

We find this simplicity to be intriguing, and suggestive of a
universality that it would be interesting to prove. The universality
itself is in line with a conjecture made some time ago about the
existence of a much simpler model which might underlie the
physics\cite{conjecture}:  Some of the gross features are similar to a
reduced dynamical model such as a large $N$ matrix model or integrable
system related to the Calogero--Moser model at large $N$.  We discuss
some of these ideas and features in section~\ref{sec:discussion}.

\section{The Ten Dimensional Geometry}

\label{sec:solution}

We first present the complete solution of ref.\cite{Pilch:2003jg}.
The Einstein frame metric is:
\begin{eqnarray*}
ds^2&=&\Omega^2(k^2\eta_{\mu\nu}dx^\mu dx^\nu)\\
&&+\,\,\Omega^{-2}\left\{H_1\left[du^2+u^2(\sigma_2^2+\sigma_3^2)\right]+H_1^{-1}u^2\sigma_1^2
+H_2dv^2+H_2^{-1}v^2d\phi^2\right\}\ ,
\end{eqnarray*}
where $k$ is a constant, and
\begin{eqnarray}
\eta_{\mu\nu}dx^\mu dx^\nu&=&-(dx^0)^2+(dx^1)^2+(dx^2)^2+(dx^3)^2\ ,\nonumber\\
H_1(u,v)&=&\frac{1}{\cos\beta}\ ,\qquad H_2(u,v)= \frac{1}{c \cos\beta}\
,\nonumber\\ \Omega(u,v)&=&\frac{u^{1/2}}{(H_1-H_1^{-1})^{1/4}}\
,
  \labell{tendee}
\end{eqnarray}
where $d\sigma_1=2\sigma_2\wedge d\sigma_3$ (and cyclic permutations)
define the left invariant Maurer--Cartan forms on $S^3$, and note
that:
\begin{equation}
H_1H_2^{-1}=c \ , \qquad H_1H_2=\partial_v(vc)\ .
  \labell{relation}
\end{equation}
The other supergravity fields of relevance here are the
axion--dilaton fields and the R--R four--form potential. These are
given as:
\begin{eqnarray}
\tau&\equiv& C_{(0)}+ie^{-\Phi}=
\frac{\tau_0-{\bar\tau}_0B}{1-B}\ ;\qquad
B=\left(\frac{1-H_2}{1+H_2}\right)e^{2i\phi}\ , \nonumber\\
C_{(4)}&=&w(u,v) dx^0\wedge dx^1\wedge dx^2\wedge dx^3\ ,\qquad
w(u,v)=\frac{k^4}{4g_s}\frac{u^2}{(H_1^2-1)}=\frac{k^4}{4g_s}\Omega^4\cos\beta\ ,
  \labell{morefields}
\end{eqnarray}
where by setting
\begin{equation}
  \label{backset}
  \tau_0=\frac{i}{g_s}+\frac{\theta_s}{2\pi}\ ,
\end{equation}
we have set the asymptotic value of the dilaton and the R--R scalar
$C_{(0)}$ in terms of the string coupling $g_s$ and the constant
$\theta_s$. These in turn set the asymptotic Yang--Mills coupling
($g^2_{\rm YM}=2\pi g_s$) and the theta angle in the $SU(N)$ gauge
theory on the D3--branes to which this geometry is holographically dual at
large $N$.

We also note here that the solution has been presented with the
natural length scale, $L$, of the spacetime, set to unity. It is given
in terms of the string tension, string coupling, and  $N$ (the
number of branes sourcing the geometry) as $L^4=4\pi(\alpha^\prime)^2 g_s N$.
This can easily be restored as needed.

For completeness, we also note that the three--form field strength
$G_{(3)}$  is given by:
\begin{equation}
  \labell{threeform}
  G_{(3)}=(1-BB^*)^{-1}(dA_{(2)}-BdA^*_{(2)})\ ,
\end{equation}
where
\begin{equation}
  \labell{twoform}
  A_{(2)}=e^{i\phi}\left(a_1 dv\wedge \sigma^1+a_2 \sigma^2\wedge
  \sigma^3+a_3\sigma^1\wedge d\phi\right)\ ,
\end{equation}
with
\begin{equation}
a_1(u,v)=\frac{i}{c}\ ,\qquad a_2(u,v)=i\frac{v}{v\partial_vc+c}\ ,\quad
a_3(u,v)=-uv\partial_u c + 2vc \ .  \labell{littleAs}
\end{equation}
As we shall see, we will not need these fields in the study presented here.

The remarkable thing about this large class of solutions is that it is
seeded entirely by the function $c(u,v)$, which is obtained as a solution
to the following non--linear differential equation:
\begin{equation}
  \labell{thediffy}
 \frac{\partial}{\partial u}\left(\frac{v^3}{u}\frac{\partial
 c}{\partial u}\right)+\frac{\partial}{\partial
 v}\left(\frac{v^3}{u}c\frac{\partial c}{\partial v}\right)=0\ .
\end{equation}

This equation is very difficult to solve exactly, and only two classes
of exact solutions (those which we study here) are known at present.
It is interesting to note, however, that a perturbative study can
reveal some structures that  might be of use for either searching for
exact solutions or for seeding numerical studies. The point is that
(as suggested in ref.\cite{Pilch:2003jg}) one can write
\begin{displaymath}
  c(u,v)=1+\sum_i c_i(u,v)\lambda^i
\end{displaymath}
and attempt to determine the functions $c_i(u,v)$ from the resulting
linearised equations. This is an interesting line of attack that we
have not carried out in great detail so far. At first order we have
noticed that if we separate variables according to
$c_1(u,v)=U_1(u)V_1(v)$, then ${\tilde U}_1=U_1/u$ and ${\tilde
  V}_1=V_1v$ both satisfy relations of the form of Bessel's equation.
This may be a clue for further study.

\section{The General Probe Result}
\label{sec:proberesult}

In the time--honoured fashion, we will probe the supergravity solution
of the previous section with a D3--brane, whose world--volume action
is:
\begin{equation}
S=-\tau_3\int_{\cal M} \!\! d^4\xi\,\,
\det\!{}^{1/2}\!\left[G_{ab}+e^{-\Phi/2}{\cal F}_{ab}\right]+\mu_3\int_{\cal M} C_{(4)}\ ,
  \labell{action}
\end{equation}
where
\begin{displaymath}
  {\cal F}_{ab}=B_{ab}+2\pi\alpha^\prime F_{ab} \ , \qquad \mu_3=\tau_3
  g_s=(2\pi)^{-3}(\alpha^\prime)^{-2}\ ,
\end{displaymath}
and the spacetime fields are pulled back \via\ the map $x^\mu(\xi^a)$
according to \eg:
\begin{displaymath}
  G_{ab}\equiv \frac{\partial x^\mu}{\partial\xi^a}\frac{\partial x^\nu}{\partial\xi^b}G_{\mu\nu}\ .
\end{displaymath}
The D3--brane will be chosen as lying in the directions
$\{x^0,x^1,x^2,x^3\}$, and so the ``static gauge'' will be chosen so
as to respectively align the world--volume coordinates
$\{\xi^0,\xi^1,\xi^2,\xi^3\}$ with the spacetime coordinates, and the
remaining transverse coordinates, denoted generically $x^i$, will be
taken to be functions, $x^i(t)$, of $x^0=\xi^0\equiv t$, allowing for
the brane's motion.  

We keep only terms quadratic in all velocities, looking to stay in the
BPS limit in the directions where this is possible. This computation
is quite standard\cite{Johnson:gi}, and so we just state the result
here with no further elaboration. We obtain an effective Lagrangian
for a point particle moving in the six transverse coordinates ${\cal
  L}=T-V$, where:
\begin{eqnarray}
T&=&\frac{\mu_3 k^2}{2g_s}\left\{H_1\left[{\dot u}^2+u^2({\dot
      \sigma}_2^2+{\dot \sigma}_3^2)\right]+H_1^{-1}u^2{\dot \sigma}_1^2
+H_2{\dot v}^2+H_2^{-1}v^2{\dot \phi}^2\right\}\ ,\nonumber\\
V&=&\frac{\mu_3 k^4}{g_s}\Omega^4\left(1-\cos\beta\right)=\frac{\mu_3 k^4}{g_s}\left(\frac{u^2}{H_1+1}\right)\ .
  \labell{proberesult}
\end{eqnarray}

\section{Specialising to the Previous Results}
\label{sec:specialize}
\begin{figure}[ht]
  \centering
  \resizebox{3.5in}{3in}{\includegraphics[angle=0]{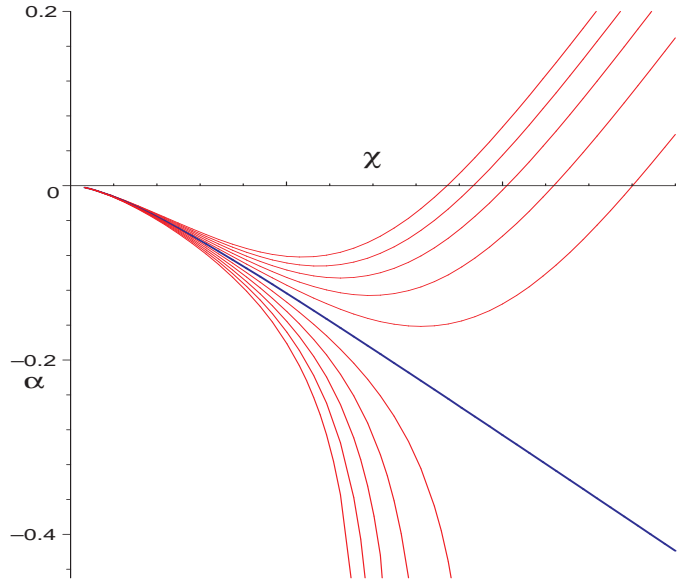}}
\caption{\footnotesize
  A family of curves of $(\alpha,\chi)$. Note that in the text, the
  natural functions discussed are $(\rho=e^\alpha, c=\cosh(2\chi))$.
  There are three natural classes, $\gamma<0$ (lying below the central curve), $\gamma=0$ (the central curve),  and
  $\gamma>0$ (above the central curve).}
\label{families}
\end{figure}

It is instructive to first obtain a few known results. The solution of
ref.\cite{Pilch:2000ue} can be recovered by choosing new coordinates
$(r,\theta)$, and a function $\rho$ such that
\begin{equation}
  \labell{special}
u(r,\theta)=\frac{\rho^3\cos\theta}{(c^2-1)^{1/2}}\ ,\qquad
v(r,\theta)=\frac{\sin\theta}{(c^2-1)^{1/2}}\ ,
\end{equation}
where the function $\rho$ is related to $c$ in the following way:
\begin{equation}
  \labell{rhorelation}
  \rho^6=c + (c^2-1)\left[\gamma+\frac{1}{2}\log\left(\frac{c-1}{c+1}\right)\right]\ ,
\end{equation}
for a real number $\gamma$, which parameterises a whole family of
solutions. It is very useful to examine the behaviour of the families
by looking at figure~\ref{families}.  The functions $(\alpha,\chi)$
plotted there are related to the functions $(\rho,c)$ by
$(\rho=e^\alpha, c=\cosh(2\chi))$. The function $\chi$ determines the
mass of the adjoint hypermultiplet in the ${\cal N}{=}2$ gauge theory
that results from the soft breaking of the pure ${\cal N}{=}4$ 
theory. In fact, the mass is related to the constant $k$ in the
solution by\cite{Buchel:2000cn} $m=k$. The function $\alpha$ sets the
vacuum expectation of the remaining ${\cal N}{=}2$ complex scalar in the
gauge multiplet.
The left hand side represents the ultraviolet (UV): the ${\cal N}{=}4$
theory with these scalars ---and hence the coefficients of the
operators to which they correspond--- switched off. As we flow to the
right the scalars switch on.  This corresponds to the
mass and the vev in the gauge theory increasing  as we flow to
the infrared (IR). Notice that for $\gamma<0$, the cases running to
$\rho=0$ all define a particular finite value of $c$, which we shall
denote $c_0$. For $\gamma=0$ this value of $c_0$ diverges.

Equations~\reef{special} give
\begin{equation}
  \labell{eliminateone}
  \rho^6=\frac{u^2(c^2-1)}{1-v^2(c^2-1)}\ ,
\end{equation}
which can be used to eliminate $\rho$. The resulting equation, after
differentiating with respect to $u$ and to $v$, removing $\gamma$,
gives  a pair of equations for $\partial_uc$ and $\partial_vc$ that  imply
the non--linear differential equation~\reef{thediffy}.

The function $\cos\beta$ can be written as:
\begin{equation}
  \labell{cosbeta}
  \cos\beta=\frac{X_1^{1/2}}{c^{1/2}X_2^{1/2}}\ ,
\end{equation}
where 
\begin{equation}
  \labell{xs}
  X_1=\cos^2\theta+\rho^6c\sin^2\theta\ ,\qquad
  X_2=c\cos^2\theta+\rho^6\sin^2\theta\ ,
\end{equation}
and 
\begin{equation}
  \labell{omegaspecial}
  \Omega^4=\frac{c^{1/2}\rho^6 X_1^{1/2}X_2^{1/2}}{(c^2-1)^2}\ ,
\end{equation}
and so the potential becomes
\begin{equation}
  \labell{potentialspecial}
  V=\frac{\mu_3 k^4}{g_s}\frac{\rho^6}{(c^2-1)^2}\left[(cX_1X_2)^{1/2}-X_1\right]\ ,
\end{equation}
and so there are two separate branches to the moduli space, where the
potential vanishes: either $\rho=0$ or $cX_2=X_1$. The latter
condition is simply $\theta=\pi/2$.

In both branches the moduli space is two dimensional, as appropriate
to the fact that we are looking at the moduli space of a single
complex scalar component that breaks the $SU(N)$ to $SU(N-1)\times
U(1)$, achieved by pulling a single brane away from the collection of
the other branes. For the first branch (which only exists for
$\gamma<0$) the coordinates are $(\theta,\phi)$ and the metric on that
space is\cite{Buchel:2000cn}
\begin{equation}
  \labell{metricone}
  ds^2_{1}=\frac{\mu_3 k^2}{2g_s}\frac{1}{c_0^2-1}\left(\cos^2\theta
  d\theta^2+\sin\theta^2 d\phi^2\right)\ .
\end{equation}
After changing variables \via\ $\sin\theta=r$, this can also be written as:
\begin{equation}
  \labell{metriconeone}
  ds^2_{1}=\frac{\mu_3 k^2}{2g_s}\frac{1}{c_0^2-1}\left(  dr^2+r^2 d\phi^2\right)\ .
\end{equation}
The parameter $r$'s maximum value is unity, marking the edge of a
disc, which is where $\theta=\pi/2$. This edge precisely matches onto
the second branch for which $\theta=\pi/2$ everywhere. The natural
coordinates on this other branch are $(c,\phi)$ and the metric
is\cite{Buchel:2000cn,Evans:2000ct}:
\begin{equation}
  \labell{metrictwo}
  ds^2_{2}=\frac{\mu_3 k^2}{2g_s}\frac{c}{c^2-1}\left(\frac{dc^2}{(c^2-1)^2}+d\phi^2\right)\ .
\end{equation}
Our interest is the vanishing of the metric, corresponding to where
singularities on the Coulomb branch appear. This occurs when $c$
diverges, which happens for $\gamma\geq 0$.

The key piece of physics to identify is the location of this singular
behaviour in terms of variables that are correctly adapted to the
${\cal N}{=}2$ gauge theory discussion\cite{Buchel:2000cn}. To find
these it is natural\cite{Buchel:2000cn,Evans:2000ct} to first find
coordinates $z=ve^{-i\phi}$ that  make the metric conformal to flat
space $dzd{\bar z}$:
\begin{equation}
  \labell{isotropic}
  \frac{dc^2}{(c^2-1)^2}=\frac{dv^2}{v^2}\qquad\Longrightarrow\qquad
  v=\sqrt{\frac{c+1}{c-1}}\ .
\end{equation}
The metric is now:
\begin{equation}
  \labell{roundstill}
  ds^2=\frac{\mu_3 k^2}{2g_s}\frac{c}{(c+1)^2}dzd{\bar z}\ .
\end{equation}
Next, we must find a coordinate change to a new coordinate $Y$ such that the metric is:
\begin{equation}
  \labell{canonical}
  ds^2=\frac{\mu_3 k^2}{2}e^{-\Phi} dYd{\bar Y}\ .
\end{equation}

The coordinate $Y$ is to represent the vacuum expectation value of the
scalar in the gauge multiplet whose moduli space we are examining. The
prefactor is the same as the one which appears in front of the kinetic
term for the gauge field, and so determines the appropriate
coordinates to use. The dilaton can be determined from equations~\reef{morefields} to be:
\begin{equation}
  \labell{dilaton}
  e^{-\Phi}=\frac{c}{g_s|\cos\phi+ic\sin\phi|^2}\ .
\end{equation}
Writing
\begin{equation}
  \labell{jacob}
  dzd{\bar z}=dYd{\bar Y} \frac{\partial z}{\partial Y}\frac{\partial{\bar z}}{\partial {\bar Y}}\ ,
\end{equation}
we have
\begin{equation}
  \labell{themess}
  \left|\frac{\partial Y}{\partial z}\right|^2= k^2\left|\frac{\cos\phi+ic\sin\phi}{c+1}\right|^2 = 
\frac{k^2}{4}\left|1-\frac{1}{z^2}\right|^2\ ,
\end{equation}
where in the last line we have used  repeatedly that $z=ve^{-i\phi}$ and that  $v=\sqrt{{(c+1)}/{(c-1)}}$.
So finally we have the elegant result:
\begin{equation}
  \labell{thechange}
  Y=\frac{k}{2}\left(z+\frac{1}{z}\right)\ .
\end{equation}
It is in terms of the quantity $Y$ we should look for the non--trivial
behaviour of the coupling. To that end, we compute $\tau(Y)$.  Using
that $\cos\beta=1$ (from equation~\reef{cosbeta}), we have from
equations~\reef{tendee} and~\reef{morefields} that $B=z^{-2}$, and hence
(setting $k=m$):
\begin{equation}
  \labell{tau}
  \tau=\left(\frac{\tau_0-{\bar\tau}_0 z^2}{1-z^2}\right)
\quad = \quad \frac{i}{g_s}\left(\frac{Y^2}{Y^2-m^2}\right)^{1/2}+\frac{\theta_s}{2\pi}\ .
\end{equation}
So finally we conclude that there is a singular locus of points where
the gauge coupling diverges on the complex $Y$ plane. It is given by
the location of the square root branch cut in the function~\reef{tau},
and is the large $N$ analogue of the Seiberg--Witten singular locus
for this particular branch of the ${\cal N}{=}2$ supersymmetric $SU(N)$
gauge theory with massive adjoint hypermultiplet of mass $m$. We will
discuss this further in section~\ref{sec:discussion}.

\section{Another Example} 
\label{anotherexample}
In ref.\cite{Pilch:2003jg}, where the new, more general solution
(displayed in section~\ref{sec:solution}) was presented, a very simple
solution to the differential equation~\reef{thediffy} that seeds the
solution was discovered. This represents a new slice of the Coulomb
branch, and we should use the techniques we have been studying to
examine it. We can simply specialise our results for the probe
computation, and attempt to carry out some of the analysis of the
previous section.

The solution is described as ``separable'', since each half of the
non--linear differential equation~\reef{thediffy} is identically zero:
\begin{equation}
  \labell{newsolution}
  c=\mu(1+bu^2)\left(1-\frac{a}{v^2}\right)^{1/2}\ ,
\end{equation}
for $a,b,\mu$ arbitrary real constants, which gives:
\begin{equation}
  \labell{Hs}
  H_1=\mu(1+bu^2)\ , \qquad H_2=\frac{v}{(v^2-a)^{1/2}}\ .
\end{equation}
After  substitution,  the potential is:
\begin{equation}
  \labell{newpotential}
  V=\frac{\mu_3 k^4}{g_s} \left(\frac{u^2}{\mu(1+bu^2)+1}\right)
\end{equation}
Now, matching to the asymptotic value of $c$ requires that $\mu=1$ and
$b=0$. Inserting these (the first is enough in fact) shows that there
is only one branch for the moduli space, which is $u=0$. The moduli
space is again two dimensional, as expected, and the metric is:
\begin{equation}
  \labell{newmodulispace}
  ds^2=\frac{\mu_3 k^2}{2g_s}(v^2\pm a)^{1/2}v\left(\frac{dv^2}{v^2\pm a}+d\phi^2\right)\ ,
\end{equation}
where we have taken $a$ to mean its positive part and hence written
the two choices of sign we can have in the metric. 

We must now find the natural ${\cal N}{=}2$ coordinates. First we write,
taking the plus sign (we will deal with the minus case later):
\begin{equation}
  \labell{newcoordstry}
  ds^2=\frac{\mu_3 k^2}{2g_s}(v^2+ a)^{1/2}v\left(\frac{dw^2}{w^2}+d\phi^2\right)\ ,
\end{equation}
for some new radial coordinate we must find, $w$. After some algebra, we find:
\begin{equation}
  \labell{newradial}
  v=\frac{a^{1/2}}{2}\left(w-\frac{1}{w}\right)\ ,
\end{equation}
and so noting the useful relations
\begin{equation}
  \labell{relations}
  v^2=\frac{a}{4w^2}(w^2-1)^2\ , \quad v^2+a=\frac{a}{4w^2}(w^2+1)^2\ ,
\end{equation}
our metric is:
\begin{equation}
  \labell{newmetrictryone}
  ds^2=\frac{\mu_3 k^2}{2g_s}\frac{(v^2+ a)^{1/2}v}{w^2}(dw^2+w^2 d\phi^2)=\frac{\mu_3 k^2}{2g_s}\frac{a}{4}\left(1-\frac{1}{w^4}\right)dzd{\bar z}\ ,
\end{equation}
where $z=we^{-i\phi}$.

Again, we must match this to the form given in
equation~\reef{canonical}. This time, we can read off the dilaton from
the supergravity solution as:
\begin{equation}
  \labell{newdilaton}
  e^{-\Phi}=\frac{H_2}{g_s|H_2\cos\phi+i\sin\phi|^2}\ .
\end{equation}
Our crucial equation is now:
\begin{equation}
  \left|\frac{\partial Y}{\partial z}\right|^2= 
\frac{ak^2}{4}\left((w^2-1)^2\cos^2\phi+(w^2+1)^2\sin^2\phi\right) 
= \frac{ak^2}{4}\left|1-\frac{1}{z^2}\right|^2\ ,
  \labell{changeitorelse}
\end{equation}
where we have used repeatedly that $z=we^{-i\phi}$. It is pleasing that
there is a such a simple result, and interesting that although the
details seem very different from the  example in the previous section, it gives precisely
the same form for the change of variables:
\begin{equation}
  \labell{thenewchange}
  Y=\frac{a^{1/2}k}{2}\left(z+\frac{1}{z}\right)\ .
\end{equation}
Since $H_2$ can be written simply as $H_2=(w^2-1)/(w^2+1)$, once again
the algebra simplifies marvellously, and we get the result that
$B=z^{-2}$; and therefore the result for the gauge coupling in the
natural ${\cal N}{=}2$ adapted complex $Y$ plane is:
\begin{equation}
  \labell{tauagain}
  \tau=\left(\frac{\tau_0-{\bar \tau}_0z^2}{1-z^2}\right)\quad=\quad\frac{i}{g_s}
\left(\frac{Y^2}{Y^2\mp ak^2}\right)^{1/2}+\frac{\theta_s}{2\pi}\ .
\end{equation}
We have recovered the possibility of the other sign for $a$. We see
that it naturally connects onto the plus sign case as follows: There
is again a square root branch cut of width set now by $a^{1/2}k$, and
it lies on the real axis for the plus sign choice (corresponding to
the minus in the expression immediately above). As $a$ goes to zero,
$c$ becomes constant everywhere, and we return to the boring ${\cal
  N}{=}4$ result.  This can be seen by examining the solution for $c$
given in expression~\reef{newsolution} (with, recall, $\mu=1$ and
$u=0$ to be on the moduli space, or alternatively $b=0$ to be
asymptotically constant). On the $Y$--plane this corresponds to the
cut shrinking to zero size and disappearing, with $\tau$ becoming a
constant, $\tau_0$. As $a$ continues to the other sign however, the
cut simply reappears, but aligned along the imaginary axis.

\section{Discussion}
\label{sec:discussion}

It is intriguing that the results of the previous section for the new
solution also give such a clean outcome in the natural ${\cal N}{=}2$
coordinates.  There is the same form as in
section~\ref{sec:specialize} for the singular locus where the gauge
coupling $\tau$ diverges. This is a special piece of the moduli space
of the full gauge theory. Generically, it is to be expected that there
are of order $N$ singular points on the Coulomb branch, and there
ought to be nothing special about their arrangement for any $N$.
Gravity and string duals of the gauge theory at large $N$ suggest the
existence of very special large $N$ limits where these singular
points coalesce into a one--dimensional locus. This locus has a
stringy and supergravity understanding as the ``enhan\c con'' locus of
ref.\cite{Johnson:1999qt}. It is associated with the place where
D3--branes (in this example), which would be pointlike in the relevant
transverse space, become tensionless and smear out transversely to
fill out a one--dimensional (in this example) submanifold densely.

In gauge theory terms, the large $N$ theory is a special limit in
which, away from the singular points, the generic instanton
corrections to the form of the coupling are suppressed (due to $N$
being large); but then they switch on strongly at short separation,
spreading the points into the singular
locus\cite{Johnson:1999qt,Buchel:2000cn}.

As we have seen from the probe results, once we get to the right
${\cal N}{=}2$ adapted coordinates the loci seem to be controlled by a
very simple structure.  Although we have only seen two classes of
exact solution for which this can be demonstrated explicitly, they
seem so different in the details but yet yield so similar a final form
that it is natural to conjecture that this simple structure will
persist: The (inverse) square root branch cut form will perhaps always
control the location of the locus in this large class of examples
given by the supergravity solution in section~\ref{sec:solution}.
 
We expect that what will distinguish the details on the Coulomb branch
is probably only two features: {\it (1)} The number of distinct such
cuts or segments that can appear in the plane ---we can imagine
multi--cut situations--- and {\it (2)} the detailed distribution of
smeared D3--branes within each cut. This will be determined by
functions $\rho_i(Y)$ that will give the D3--brane density in the
$i$th segment on the complex $Y$--plane. These details are implicit in
the precise functional dependence of $c(u,v)$, which translates into a
specific relation between the mass and the choice of the precise
pattern of vacuum expectation values breaking $SU(N)$ to
$SU(N-1)\times U(1)$ (corresponding in the dual theory to pulling off
a single D3--brane from the group of $N$).  Perhaps a study along the
lines of ref.\cite{Buchel:2000cn} could be carried out for new
examples to determine the density distributions. Notice that, in the
example reviewed in section~\ref{sec:specialize}, the density
function extracted in ref.\cite{Buchel:2000cn} was $\rho(Y)\sim
\sqrt{m^2-Y^2}$, which is in fact a semi--circle.

There are a number of suggestive features here, which support a
conjecture made some time ago\cite{conjecture} about the presumed
underlying simplicity of the physics in question. The conjecture is
that there is a reduced model appearing at large~$N$ that controls the
physics: There may be a matrix model (or closely related
integrable model) of some variety responsible for these broad
features. Such models at large $N$ are known to have exactly the
attributes required: {\it (1)} They have simple distributions of
eigenvalues or charges, given by density functions $\rho_i(Y)$ (often
of Wigner's semi--circle type), sometimes with support on a number of
disconnected segments (labelled $i$ here) described in terms of square
root branch cuts in the plane.  {\it (2)} Many details of any
additional potentials these models might have are irrelevant at
strictly large $N$, and so they often fall into simple universality
classes.  The same universality may be at work here. {\it (3)} Such
models are often associated with non--linear differential equations.
Perhaps the equation~\reef{thediffy}, seeding the entire class of
supergravity solutions, may have its origins in the context of these
simpler models.

The matrix model is expected to be closely related to the
Calogero--Moser system. The reasons for this expectation are
circumstantial, but worth mentioning. The point is that, at any $N$,
the Calogero--Moser model has been shown to share the same formal data
as the Seiberg--Witten solution for the associated $SU(N)$ gauge
theory\cite{D'Hoker:1999ft}.  In particular, the Seiberg--Witten curve
is essentially the spectral curve of the integrable system defined by
the Calogero--Moser model.  The Seiberg--Witten curve of course
encodes the physics of the Coulomb branch of the gauge theory and, in
particular, its points of degeneration give the places where the gauge
coupling diverges. The idea\cite{conjecture} would be that at large
$N$ this relation becomes more than formal: the dynamics may have
their {\it best} description in terms of the variables of the
integrable model. It is also of note that the Calogero--Moser model
(in some limits) can be derived from a simple matrix
model\cite{Polychronakos:1997nz}, and that at large $N$ the
distribution of the interacting charges (the eigenvalues) in the model
is again Wigner's semi--circle distribution on the line\cite{distrib}.
The description of the $1/N$ corrections may then have their home in
familiar matrix model and integrable system technology (see also
ref.\cite{Ferrari:2001zb} for related work which may support some of
these ideas).

This is very natural also from the point of view of branes, of course.
Branes have come to be recognised as having more than a passing
resemblance to eigenvalues in some type of matrix model, and this
context would be one way to make that precise. This was part of the
motivation for the conjecture in ref.\cite{conjecture}; the idea was
to find effective models of the enhan\c con locus as a dynamical
object in its own right. The constituent branes have become
tensionless and, furthermore, $N$ of them have coalesced into a single
unit and so their description in the usual terms is difficult. The
expectation was that the sought--after matrix model would be a new
collective description of the enhan\c con. This hope remains. Although
the idea seemed somewhat far--fetched at the time, the results
presented here together with the new excitement about matrix models'
relevance to four dimensional gauge theory\cite{Dijkgraaf:2002fc}
suggest that there may be hope to find such a model.

\section*{Acknowledgements}
JEC is supported by an EPSRC  studentship. CVJ's research is
supported in part by the EPSRC.  This manuscript is Durham CPT
preprint number DCTP-03/25.

\end{document}